\documentclass[a4paper]{aa}  

\usepackage{amsmath}
\usepackage{graphicx}
\usepackage{longtable}
\usepackage{txfonts}
\usepackage{natbib}
\usepackage{amssymb}
\usepackage{mathrsfs}
\usepackage{stmaryrd}

\providecommand{\abs}[1]{\lvert#1\rvert}
\begin{document}
\setcounter{table}{0}
\setcounter{figure}{0}

\title{Prospecting transit duration variations \\
 in extrasolar planetary systems}
\titlerunning{Prospecting TDVs}
    \authorrunning{C. Damiani \& A. F. Lanza}

%   \subtitle{}

   \author{C.~Damiani \and A.~F.~Lanza}

   \offprints{C.~Damiani}

   \institute{INAF-Osservatorio Astrofisico di Catania, Via S. Sofia, 78
               -- 95123 Catania, Italy \\
              \email{damiani@oact.inaf.it} }

   \date{Received ... ; accepted ... }

%%%%%%%%%%%%%%%%%%%%%%%%%%%%%%%%%%%%%%%%%%%%%%%%%%%%%%%%%%%%%%%%%%%%%%%

\abstract{Transiting  planetary systems allow us to extract geometrical information, e.g., the angle $\epsilon$ between the orbital angular momentum and the stellar spin, that can be used to discriminate among different formation and evolutionary scenarios. This angle is constrained by means of the Rossiter-McLaughlin effect observed on radial velocity and can be subject to large uncertainties, especially for hot stars ($T_{\rm eff} > 6250$~K). It is thus interesting to have an alternative method to constrain the value of the obliquity $\epsilon$ and to detect companions that might have disturbed the orbit of the planet.
}{We show how the long-term variations in the transit duration (TDV) can be used to constrain the obliquity of the stellar rotation axis. Our calculations may also be used to put an upper limit on the contribution of geometrical effects to the TDVs, thus allowing us to indirectly infer the presence of additional companions.
}{We introduce a simple theory to describe the secular variations in the orbital elements and  their effects on the TDVs with a general formulation valid for both oblique and eccentric systems. Parameters or orbital elements that cannot be directly measured, such as the longitude of the ascending node of the orbit, are  avoided thus allowing us to perform a straightforward application.}{We compute the expected TDVs for the presently known transiting systems, adopting  their  parameters found in the literature. Considering the capabilities of the present or next generation space-borne telescopes, we point out the systems that could be readily observed and discuss the constraints derivable on their  fundamental parameters. }{Measured TDVs can be used to constrain the obliquity of the stars (and possibly  of the planets in systems younger than 10 -- 100 Myr), giving information about the formation scenarios, the strength of the tidal coupling, and  the internal structure of both the stars and the planets. Moreover, they can  provide an indirect indication of other bodies, even with a mass comparable with that of the Earth, because they give rise to  additional contributions to the nodal precession.}
\keywords{Planetary systems -- stars: rotation -- planet-star interactions }

%%%%%%%%%%%%%%%%%%%%%%%%%%%%%%%%%%%%%%%%%%%%%%%%%%%%%%%%%%%%%%%%%%%%%%%%
   \maketitle

%________________________________________________________________

\section{Introduction}
\label{intro}

Planetary systems with transiting planets are a fundamental source of information on stellar and planetary properties, in particular they allow us to measure the mass and the radius of the planets by removing the inclination degeneracy that affects systems without transits. Moreover, the angle $\epsilon$ between the orbital angular momentum  and the stellar spin axis can be measured in such systems and enable us to discriminate among different formation and evolutionary scenarios. In the case of hot Jupiters, if the planet underwent inward migration by angular momentum exchange with a protoplanetary disc \citep{Linetal96}, the stellar spin should be almost aligned with the orbital angular momentum, whereas if the small semi-major axis  of its orbit were due to the Kozai mechanism \citep{Kozai1962} or planet-planet scattering \citep{RasioandFord96}, high obliquities would generally be expected. If the planet transits across the stellar disc, the radial velocity anomaly detected during the transit, i.e., the so-called  Rossiter-McLaughlin effect \citep[see e.g.][and references therein]{Triaud2010} can be used to measure the projected obliquity $\lambda$, i.e., the angle between the projections on the plane of the sky of the orbital angular momentum and the stellar spin. From the value of $\lambda$, it is possible to derive the obliquity $\epsilon$ if the inclination $i_{\rm s}$ of the stellar spin to the line of sight is known (cf. Eq.~\ref{eps_equa} of Sect.~\ref{theory2.1}). 

Since it has been empirically suggested that hot stars ($T_{\rm eff} > 6250$~K) with hot Jupiters have high obliquities \citep{Winnetal10}, it is very interesting to apply different methods both to constrain the obliquity of transiting systems in general and to look for the presence of additional companions that may have perturbed the orbit of the planet. 

In this work, we illustrate how the long-term variation in the transit duration (hereafter TDV) can be used to constrain the obliquity of the stellar rotation axis and to derive indirect evidence of other bodies in a planetary system. We study the general case of a system with both an oblique and eccentric orbit giving an expression for the TDV that contains only measurable quantities. This is useful for a straightforward application of our results to the known transiting systems and to apply the method to ground-based and space-borne transit surveying and monitoring. 

\section{Apsidal and nodal precessions}

Perturbations of the Keplerian orbit occur whenever the force field controlling the motion of the planet is not spherically symmetric or does not follow the inverse-square dependence on the orbital radius.  \citet{Miralda02}  considered two cases: the precession of the orbital plane of a circular orbit in a misaligned system  and the precession of the periastron when the orbit is eccentric, in the limit of a small eccentricity and assuming that the orbital plane coincides with the stellar equator. In both cases, the cause of the precession is the gravitational quadrupole moment of the star. 
The latter case was further extended to an arbitrarily large eccentricity  by \citet{Pal2008},  independently of the physical mechanism causing the precession of the periastron. \citet{Iorio2011} then proposed  a set of uniform self-consistent analytic expressions accounting simultaneously for the effects of all the perturbing accelerations on the duration and the mid-time of the transit. 

In the present work, we identify the known systems for which a TDV can be easily observed over a time interval of the order of a decade, so that they can be readily used to constrain the spin-orbit alignment. Moreover, the apsidal precession in systems with an eccentric orbit may be used to estimate the Love number of the planet that is related to its internal density stratification, as shown by \citet{Ragozzine2009}. 

We introduce a simple approach, using the formulations of \citet{MardlingLin02} for the secular variations in the orbital elements,  merging the results of \citet{Miralda02} and \citet{Pal2008}, to assess their effects on TDVs. In our formulation, parameters or orbital elements  that cannot be directly measured, such as the longitude of the ascending node of the orbit, are avoided, allowing us a direct application of our theory.  

\subsection{Rate of change in the orbital elements}
\label{theory2.1}

We assume a cartesian orthogonal reference frame $\mathcal{R}_0$ with basis vectors $({\mathbf{{i}}}, \mathbf{{j}}, \mathbf{{k}})$,  the origin at the barycentre of the host star, and the $\mathbf{{k}}$ vector along the line of sight, so that the $(\mathbf{{i}}, \mathbf{{j}})$ plane coincides with the plane of the sky. 
Following \citet{MardlingLin02}, the  orbital elements are expressed through the specific relative angular momentum vector $\mathbf{h = r} \times \mathbf{\dot{r}}$, where $\mathbf{r}$ is the position vector of the barycentre of the planet,  and the Runge-Lenz vector $\mathbf{e}$, a vector in the direction of the periastron with magnitude equal to the eccentricity of the orbit. By definition, the vectors $\mathbf{h}$ and $\mathbf{e}$ are always perpendicular, although their components in the adopted reference frame will change in time owing to the perturbing potential. The secular evolution in the elements is obtained by time-averaging over the orbit, using the true anomaly as the integration variable. The eccentricity $e$, the semimajor axis $a$, the inclination $i$, the argument of periastron $\omega$, and the longitude of the ascending node $\Omega$ may be obtained from $\mathbf{h}$ and $\mathbf{e}$ through equations (21)--(27) of \citet{MardlingLin02}.

For a two-body system, the average rate of change in $\mathbf{h}$ and $\mathbf{e}$ can be written as the sum of the different perturbation contributions averaged over one orbit, i.e. 
\begin{align}
\langle\mathbf{\dot{h}}\rangle&=\langle\mathbf{d}_{\rm QD_s}\rangle + \langle\mathbf{d}_{\rm QD_p}\rangle+ \langle\mathbf{d}_{\rm TF_s}\rangle +\langle \mathbf{d}_{\rm TF_p}\rangle, \\
\langle\mathbf{\dot{e}}\rangle&=\langle\mathbf{g}_{\rm QD_s}\rangle + \langle\mathbf{g}_{\rm QD_p}\rangle+ \langle\mathbf{g}_{\rm TF_s}\rangle + \langle\mathbf{g}_{\rm TF_p}\rangle + \langle\mathbf{g}_{\rm GR}\rangle, 
\end{align}
where the different subscripts denote the perturbations as follows:
\begin{list}{--}{}
\item  ${\rm QD_i}$ is the effect of the quadrupolar distortion of the star, for $i=s$, and of the planet, for $i=p$, respectively; the quadrupole moment of each body arises from its spin centrifugal distortion as well as  from the tidal distortion caused by the other body;
\item ${\rm TF_i}$ is the effect of the tidal damping within the star, for $i=s$, and within the planet, for $i=p$, respectively;
\item ${\rm GR}$ is the effect of the (gravitoelectric) post-Newtonian potential\footnote{We neglect the usually much smaller gravitomagnetic effect, i.e., the Lense-Thirring precession \citep[see ][]{Iorio2011}.}
\begin{equation}
\langle\mathbf{g}_{\rm GR}\rangle=\frac{3 a^2 n^3}{c^2} \frac{e}{1-e^2} \, \mathbf{\hat{q}},
\end{equation}
where $c$ is the speed of light, $\mathbf{\hat{q}} \equiv \mathbf{\hat{h}} \times\mathbf{\hat{e}}$  is the unit vector orthogonal to the unit vectors in the directions of $\mathbf{h}$ and $\mathbf{e}$, respectively, and $n$ is the mean motion
\begin{equation}
n^2 a^3= G (M_{\rm s} + M_{\rm p}),
\end{equation}
$G$ being the gravitation constant, and $M_{\rm s}$ and $M_{\rm p}$ the mass of the star and the planet, respectively.
\end{list}
We focus on the evolution of the orbital elements that affect the observables of the transit on a relatively short timescale, i.e., of the order of a decade. Consequently, the terms $\mathbf{d}_{\rm TF_s}, \mathbf{d}_{\rm TF_p}, \mathbf{g}_{\rm TF_s}$, and $ \mathbf{g}_{\rm TF_p}$ related to the tidal effects are neglected, as they induce variations with typical timescales at least four orders of magnitude longer\footnote{Only in a few systems, e.g., CoRoT-11, might the tidal interaction induce variations in the orbital elements with observable consequences over a time interval of $\sim 30 $~yr if the stellar tidal quality factor $Q^{\prime}_{\rm s}$ is smaller than $(3-5) \times 10^{6}$ \citep[see ][]{Lanza11}.}. Therefore, the angles of the stellar and planetary spins  to the total angular momentum will be considered constant, as well as the values of the orbital eccentricity and semi-major axis.

 Considering that the planetary spin evolves to a coplanar state of pseudo-synchronization with the orbit within a typical timescale of  $\tau \approx 10^{5}$~yr for a planetary  tidal quality factor $Q^{\prime}_{\rm p} \simeq 10^{6}$ \citep{Leconteetal10}, we assume that the planet is already in this state and thus the planet's angular velocity can be written as $\mathbf{W}_{\rm p} = n\,\mathbf{\hat{h}}$. 
The angular velocity of the star $\mathbf{W}_{\rm s}$ can be described in $\mathcal{R}_0$ by the angles $i_{\rm s}$ and $\phi_{\rm s}$, which are respectively the inclination to the line of sight and the azimuthal angle, so that the corresponding unit vector can be written
\begin{equation}
\mathbf{\hat{W_{\rm s}}} = \begin{bmatrix}
                      \sin i_{\rm s} \cos \phi_{\rm s}\\
                      \sin i_{\rm s} \sin \phi_{\rm s}\\
                      \cos i_{\rm s}
                     \end{bmatrix}_{\mathcal{R}_0}. 
\end{equation}
The angular velocity of the star $W_{\rm s}$ is related to its equatorial rotational velocity $\tilde{v}$ (assuming rigid rotation) by $\tilde{v} = W_{\rm s} R_{\rm s}$, where $R_{\rm s}$ is the radius of the star. The spin of the star $\mathbf{L}_{\rm s}$ is defined as $\mathbf{L}_{\rm s}\equiv \underline{C_{\rm s}} \mathbf{W}_{\rm s}$,
where $\underline{C_{\rm s}}$ is the tensor of inertia of the star.  Its trace $C_s= \alpha M_{\rm s} R_{\rm s}^2$ is the moment of inertia of the star, where $\alpha$ is the fractional gyration radius.

We define $\epsilon$ to be the angle formed by the orbital angular momentum vector and the spin of the star, viz. $\cos \epsilon = \mathbf{\hat{W}}_{\rm s} \cdot \mathbf{\hat{h}}$, and let $\lambda$ be the angle formed by the projections of the orbital angular momentum and stellar spin on the plane of the sky. There is a simple relationship between $\phi_{\rm s}$ and $\lambda$, i.e. $\phi_{\rm s} = \Omega+\lambda+ \frac{\pi}{2}$; moreover
\begin{equation}
  \cos \epsilon~=~\cos i \cos i_{\rm s} + \sin i \sin i_{\rm s} \cos \lambda. \label{eps_equa}
\end{equation}
Applying those geometric relationships and inserting \citeauthor{MardlingLin02}'s Eqs. (48) and (49) into Eqs. (42) and (43), and then using their Eqs. (29), (30), and (31), the quadrupolar distortions of the star and the planet and the general relativity effect cause an orbit-averaged variation in the orbital elements given by
\begin{eqnarray}
\langle\frac{d i}{dt} \rangle&=& -A_{\rm s} W^2_{\rm s} \cos \epsilon \sin i_{\rm s} \sin \lambda, \label{eqomedot0} \\
\langle\frac{d \Omega}{dt}\rangle&=& -A_{\rm s} W^2_{\rm s} \frac{\cos \epsilon (\cos i_{\rm s} - \cos i \cos \epsilon)}{\sin i}, \label{eqomedot1} \\
\langle\frac{d\omega}{dt}\rangle &=& A_{\rm s} \left[ \frac{W_{\rm s}^2}{2}(3 \cos^2 \epsilon -1) + \frac{G M_{\rm p}}{a^3} f_0(e) \right] \label{eqomedot2}\\
&&+ A_{\rm p} \left[ n^2 + \frac{G M_{\rm s}}{a^3} f_0(e) \right] + \frac{3 a^2 n^3}{c^2}\frac{1}{1-e^2} - 
    \langle\frac{d \Omega}{dt}\rangle \cos i,  \nonumber
\end{eqnarray}
where
\begin{equation}
f_0(e) = 15 \frac{1+\frac{3}{2}e^2+\frac{1}{8}e^4}{(1-e^2)^3}, 
\end{equation}
and $A_{\rm s}$ and $A_{\rm p}$ are factors that depend on the perturbing accelerations caused by the stellar and planet bulges, respectively, i.e.
\begin{align}
A_{\rm s} &=\frac{k_{\rm s}}{n(1-e^2)^2}\left( 1+ \frac{M_{\rm p}}{M_{\rm s}}\right)\left(\frac{R_{\rm s}}{a}\right)^5\label{asoverh},\\
A_{\rm p} &=\frac{k_{\rm p}}{n(1-e^2)^2}\left( 1+ \frac{M_{\rm s}}{M_{\rm p}}\right)\left(\frac{R_{\rm p}}{a}\right)^5\label{bpovere},
\end{align}
where $R_{\rm s}$ and $R_{\rm p}$ are the radius of the star and of the planet, respectively, $k_{\rm s}$ is the apsidal motion constant of the star,  and $k_{\rm p}=\frac{1}{2} k_{\rm L}$, where $k_{\rm L}$ is the Love number of the planet. For the hot Jupiters,  the Love number is expected to be in the range $k_{\rm L} \approx 0.1 - 0.6$ \citep{Ragozzine2009}, so we assume that $k_{\rm L}=0.3$. The values of $k_{\rm s}$ are computed as a function of the mass and the effective temperature of the star using the stellar models by \citet{Claret95} assuming a solar composition. For transiting hot-Jupiters, we usually find that $A_{\rm p}/A_{\rm s}~\simeq~100  $, which implies that the planetary distortion is predominant for orbital periods shorter than $\sim 3$~days, allowing a measurement of the planet Love number \citep[see ][]{Ragozzine2009}. 

For polar and equatorial orbits ($\epsilon = 90^\circ$ and $\epsilon=0^\circ$, i.e. $\lambda=0^\circ$), there is no nodal precession because the torque applied to the planetary orbit by the stellar quadrupole vanishes. In contrast, the perturbation to the inverse-square law of the force introduced by the quadrupole term of both components and the relativistic effect cause an apsidal precession for all values of the obliquity $\epsilon$.

\begin{table*}[t]
\centering
\begin{tabular}{lccccccccc}
\hline\hline
Name & $T_{\rm eff}$ (K) & $e$ & $\omega$ ($^\circ$) & $\lambda$ ($^\circ$)& V mag& Refs. & $\Delta H$ (s/yr)& $\Delta H|_{\omega}$ (s/yr) &  $\Delta H|_{i}$ (s/yr)\\
\hline
CoRoT-1    &  $5950\pm150$  & $0$               & - &  $77    \pm  11  $ &$13.6$ & 3, 12
  &  $-1.12 $   &  $ -1.12 $   &  - \\
*CoRoT-2		&  $5625\pm120$  & $0$               & - &  $-7.2   \pm  4.5  $ & $12.6$& 1, 5
  &  $0.92 $   &  $ 0.92$   &  - \\
HAT-P-2    &  $6290\pm110$  & $0.5163\pm0.0025$ & $185.22\pm0.95$ & $0.2   \pm  12.5$ &$8.71$& 17, 11
  &  $ 3.04 $   &  -   &  $3.04$\\
WASP-33     &  $7430\pm100$  & $0$               & - &  $-107.7\pm  1.6 $ & $8.3$ &6
  &  $-34.80 $   &  $ -34.80 $   &  - \\
Kepler-8   &  $6213\pm150$  & $0$     & - & $-26.9 \pm  4.6 $ &$13.89$&8, 10
  &  $2.06 $   &  $2.06$   &  - \\
TrES-4      &  $6200\pm75$   & $0$               &  - & $-6.3  \pm  4.7 $ &$11.6$&14
  &  $ 0.73 $   &  $  0.73 $   &   - \\
*WASP-3		&  $6400\pm100$ & $0$  				& - &  $3.3\pm3.45$		& $10.5$& 16 & $-1.03$ &$-1.03$ & - \\
WASP-17     &  $6550\pm100$  & $0.24\pm0.07$     &  $278\pm6.5$ & $-148.5\pm  5.1 $ & $11.6$& 2, 15
  &  $-5.82 $   &  $ -1.16 $   &  $-4.66$\\
WASP-18     &  $6400\pm100$  & $0.0085\pm0.001$  &  $-92.1\pm4.9$ & $-5.0  \pm  3.1 $ & $9.3$&7, 15
  &  $ 1.79$   &  $ 1.75 $   &  $0.04 $\\
*WASP-38		&  $6150\pm80$  & $0.0314\pm0.0046$ & $344.0\pm17.5$ & $15\pm43$ & $9.4$& 4, 13 & $-0.71$ & $-0.69$ & $-0.005$\\
XO-3        &  $6429\pm100$  & $0.2884\pm0.0035$ &   $346.3\pm1.3$ & $-37.3 \pm  3.7 $ & $9.8$&18, 9
  &  $ 12.70 $   &  $ 12.73$   &  $-0.03$\\
 \hline
\end{tabular}
\caption{Transiting planetary systems with measured misalignment for which $\dot{H} \geq 2 \times 10^{-8}$. The asterisk  denotes the systems that have a significant TDV when taking the extreme possible values for $\lambda$, see text for details. References: 1 : \citet{Alonso2008}; 2: \citet{Anderson2010}; 3: \citet{Barge2008}; 4:  \citet{Barros2011};  5: \citet{Bouchy2008}; 6: \citet{Collier2010}; 7: \citet{Hellier2009}; 8: \citet{Jenkins2010}; 9: \citet{Johns2008};  10: \citet{Kipping2011}; 11: \citet{Pal2010}; 12: \citet{Pont2010}; 13: \citet{Simpson2010}; 14: \citet{Sozzetti2009}; 15: \citet{Triaud2010}; 16: \citet{Tripathi2010}; 17:\citet{Win2007}; 18: \citet{Win2009}. }% HD147506/HAT-P-02  HD15082/WASP-33}
\label{tab1}
\end{table*}

\subsection{Variation in the transit duration}

We denote the duration of the transit by $H$, which is defined as the interval between the times when the centre of the planetary disc intersects the limb of the star during the ingress and the egress, respectively. Neglecting the curvature of the projection of the orbit because of its inclination, $H\simeq  \frac{\overline{AB}}{v_{\rm tan}}$, where $\overline{AB}$ is the chord travelled by the centre of the planet during its transit on the stellar disc with a tangential velocity $v_{\rm tan}$. We  define the angle $\gamma$ so that $\overline{AB} \equiv 2 R_{\rm s} \cos \gamma$, thus
\begin{equation}
H = \frac{2 R_{\rm s}}{v_{ \rm tan}} \cos \gamma. \label{eq:h0}
\end{equation}
The tangential velocity can be expressed as \citep{Pal2008}
\begin{equation}
v_{\rm tan}= a n \frac{1+e \cos \nu}{\sqrt{1-e^2}}, \label{eq:tanvelo}
\end{equation}
where $\nu$ is the true anomaly of the planet. We assume that the planet has a constant velocity during the transit equal to the velocity at mid-transit. The true anomaly $\nu$ and the phase angle $\theta$ are related by $\theta = \nu + \omega - \pi/2$. When $i=90^\circ$, the primary minimum occurs when $\theta = 0^{\circ}$ and $\nu = \pi/2 - \omega$. For transiting systems, $i \simeq 90^\circ$, thus we can apply these relationships without introducing an appreciable error and obtain 
\begin{equation}
v_{\rm tan}= a n \frac{1+e \sin \omega}{\sqrt{1-e^2}}\quad \text{and} \quad \dot{v}_{\rm tan}= a n \frac{e \cos \omega}{\sqrt{1-e^2}} \dot{\omega}.
\label{eq:vtan}
\end{equation}
We note that for nearly circular orbits (i.e. when the eccentricity and/or the Lagrangian orbital elements
$e\cos\omega$ and $e\sin\omega$ are zero within a few $\sigma$), the true anomaly and $\omega$ are meaningless, but the phase angle $\theta$ has a well-defined value. In this case, the tangential velocity at the time
of each transit do not change within the uncertainties.
We define $b'$ to be the projected distance between the centre of the planet and the centre of the star at mid-transit, i.e, the impact parameter\footnote{The impact parameter is often given in units of the stellar radius, although here we use the non-normalized value.}, $b' \equiv R_s \sin \gamma$. In the framework of our approximations \citep{Kopal1959}
\begin{equation}
b'=\frac{a (1-e^2) \cos i}{1+e \sin \omega}.
\label{eq:bprime}
\end{equation}
Thus differentiating Eq. \eqref{eq:h0} with respect to the time and inserting equations \eqref{eq:vtan} and \eqref{eq:bprime}, we find 
\begin{eqnarray}
\dot{H}  & = & \frac{2}{v_{\rm tan} \sqrt{R_{s}^{2} - b^{\prime 2}} } \left[ b^{\prime 2} \tan i \langle \frac{di}{dt} \rangle  + \right. \nonumber \\ 
 &  &  \left. (2 b^{\prime 2} - R_{s}^{2}) \frac{e \cos \omega} {1 + e \sin \omega} \langle \frac{d\omega}{dt} \rangle \right]. 
\label{hdot}
\end{eqnarray}

This equation is valid for all the values of $b'$ except $|b'| = R_s$, i.e. when the projected trajectory of the planet is tangential to the disc of the star and there is no transit.

\subsection{Method to observe the TDV and its accuracy}
\label{observation_method}

Equation~(\ref{hdot}) gives the TDV as a function of the variation in the orbital elements and parameters derived from the modelling of the transit light curve. On the other hand, the variations in the orbital elements depend on the orbital parameters of the system, the stellar rotation rate, and the inclination $i_{\rm s}$ of the stellar spin to the line of sight (cf. Eqs.~\ref{eqomedot0}, \ref{eqomedot1}, and \ref{eqomedot2}). The stellar rotation period can be derived from the rotational line broadening, the radius, and the inclination of the stellar spin $i_{\rm s}$, or from the rotational modulation of its optical flux. The inclination $i_{\rm s}$ can be measured in the case of a late-type star by asteroseismic methods \citep[e.g., ][]{Ballotetal06} or, for a moderately active star, by a suitable spot modelling applied to high-precision photometric time series such as those obtained with CoRoT or Kepler \citep[][]{Mosseretal09}. 

In the case of a circular orbit, the precession of the line of the nodes induces an apparent inclination change that modifies the transit duration. The first term on the right hand side of Eq. \eqref{hdot} is indeed found to be caused by nodal precession, which we denote as $\left(\frac{\partial H}{\partial t}\right)_{\omega}$ [the subscript $\omega$ indicates that the argument of the periastron is fixed during the variation], is identical to the expressions of $dt_d/dt$ found in \citet{Miralda02} for circular orbits with the advantage that we do not introduce angles that cannot be measured because our $(\mathbf{i}, \mathbf{j})$ plane is the plane of the sky and not the invariable plane of the system. The second term in Eq.~\eqref{hdot}, which we denote as 
$\left(\frac{\partial H}{\partial t}\right)_{i}$, is non-vanishing only for an eccentric orbit and is related to the  precession of the line of the apsides  \citep[cf. ][]{Pal2008}. As already noted by several authors, $\left(\frac{\partial H}{\partial t}\right)_{i}$ vanishes when the impact parameter $b'= \sqrt{2}R_{\rm s}/2$. 

The duration of the transit can be determined with an accuracy that depends on the  depth of the transit, the accuracy  of the photometry, and the presence of  noise sources, e.g., related to stellar activity. In the case of the spatial mission CoRoT, the photon-limited photometric accuracy ranges typically from $\sim 75$ ppm in a hour integration time to $\sim 1130$ ppm/hour for a star in the magnitude range $11 \leq R \leq 16$ \citep{Aigrainetal09}. The Kepler telescope can reach $\sim 30$ ppm/hour to $\sim 700$ ppm/hour over the same range of magnitudes \citep{Jenkinsetal2010a}.  One of the most precise determination of transit duration was that obtained for CoRoT-11. In this case, stellar magnetic activity has a negligible impact because the stellar optical flux does not show a detectable modulation induced by starspots. Therefore, the accuracy is dominated by the transit depth and the photometric accuracy. The depth of the transit is $\approx 0.011$ mag and the star has $V = 12.94$ giving a photometric accuracy of $\sim 250$ parts per million for measurements collected by the CoRoT telescope with a cadence of $\sim 130$~s \citep{Gandolfi2010}. The accuracy of the transit duration is $\sim 50$~s considering a sequence of $\sim 50$ consecutive transits for the modelling. Since the aperture of the CoRoT telescope is $27$~cm, a dedicated space-borne telescope with an aperture of $\sim 3$~m should easily reach an accuracy of $\sim 5$~s in the measure of $H$ by acquiring data with a shorter cadence. Considering a time interval of 10 years, i.e., $\sim 3.2 \times 10^{8}$~s, we estimate that $\dot{H} \simeq 2 \times 10^{-8}$  can be assumed as our detection limit, corresponding to a variation of the transit duration $\Delta H$ of $\sim 0.6$~s/yr. \citet{Miralda02} obtained the same limit for HST observations of the brighter target HD~209458 ($V=7.64$) over a time interval of three years. 
Hot stars ($T_{\rm eff} > 6000-6200$~K) are the most suitable candidates for this kind of measurement because of their more likely misalignment and their low level of magnetic activity, making the distortion of the transit profile and its depth variations induced by starspots negligible.

In addition to the light dip produced by the transit of the planet across the disc of its host star, it is possible to observe the occultation of the planet by the star, i.e., the secondary eclipse. This observation is easier in the infrared because of the most favourable flux ratio  and has indeed been performed for some of the systems cited in this study, e.g., CoRoT-1 \citep{Alonso2009}, CoRoT-2 \citep{Gillonetal10}, WASP-12 \citep{Croll2011}, and WASP-18 \citep{Nymeyer2010}. The secondary eclipse of OGLE-TR-56 \citep{Sing2009} has been observed in the visible from the ground.
The timing of the mid-occultation with an uncertainty of a few seconds has also allowed us to estimate $e \cos \omega$ for those systems with a precision of the order of $10^{-4}$. Moreover, the different durations of the transit and the occultation can be used to measure $e \sin \omega$, although with a lower precision \citep{Ragozzine2009}.
In Eq.~(\ref{hdot}), we note that, whereas an increasing inclination of the orbit (i.e., $di/dt > 0$) causes an increase in the duration of both the transit and the  occultation, the sign of $\left( \frac{\partial H}{\partial t} \right)_{i}$ changes in the case of the occultation because in this case $\pi$ must be added to the argument of the  periastron in Eq.~(\ref{eq:vtan}).

\subsection{Variations in the time of the mid-transits for eccentric systems}
\label{other_causes}

In the case of an eccentric orbit, the precession of the periastron will not only cause TDVs but also a variation in the time of the mid-transit (hereafter TTV) leading to a slow change in the observed orbital period $P_{\rm obs}$.
 Since we have neglected the effect of the changing inclination of the system $i$ on the  true anomaly of the planet at the time of mid-transit (cf. Eq. \ref{eq:tanvelo} and \ref{eq:vtan}), we  assume that the nodal precession will not affect the rate of change of the observed period $\dot{P}_{\rm obs}$. Therefore, we use the derivations of \citet{Pal2008}, valid for $i=90^\circ$ and arbitrary eccentricity
\begin{equation}
\dot{P}_{\rm obs}= \frac{(1-e^2)^{3/2} e \cos \omega}{(1+e \sin \omega)^3}\frac{P_0^2}{\pi} 
\left( \langle \frac{d \omega}{dt} \rangle  \right)^{2}, 
\label{pdot_peri}
\end{equation}
where $P_{0}$ is the true orbital period, i.e., $P_0=2\pi/n$. The difference $O-C$ between the observed epoch of mid transit and that computed with a constant-period ephemerides is 
\begin{equation}
O-C = \frac{1}{2} P_{0} \dot{P}_{\rm orb} N^{2},
\end{equation}
where $N$ is the number of transits elapsed from the initial epoch and we have considered a time interval much shorter than $2\pi / \langle \frac{d\omega}{dt} \rangle $ so that $\dot{P}_{\rm obs}$, given by Eq.~(\ref{pdot_peri}), is virtually constant.
We note that the sign of $\dot{P}_{\rm obs}$ is the opposite for the occultation because in this case $\pi$ must be added to the argument of the periastron $\omega$ in Eq.~(\ref{pdot_peri}). Therefore, the $O-C$ of the occultation will have the opposite sign of that of the transit, by analogy with the case of eclipsing close binaries with eccentric orbits \citep[cf., e.g., ][]{Gimenezetal87}.  

Finally, we note that the presence of a third body in the system can induce a further precession of the orbital plane leading to an additional term in $\langle \frac{d \omega}{dt} \rangle $ \citep[cf. ][]{Miralda02}, that will change the value of $\dot{H}$ and $\dot{P}_{\rm obs}$. If the semi-major axis of the third body orbit is much larger than that of the planet, the light-time effect due to the motion of the star-planet system around the barycentre of the triple system is the dominant effect and we expect to observe the same variation in the times of both the transit and the occultation. If the third body orbit is close, it is likely to be in a mean motion resonance with the transiting planet orbit, a configuration which ensures the stability of the system. In this case, TTVs of several tens of seconds or minutes are expected, even for a third body with a mass comparable to that of the Earth \citep{Agoletal05}. Moreover, radial velocity measurements can be used to look for  a close third body, allowing us to test this interpretation.

In conclusion, the observation of a TDV without an accompanying TTV can be considered as an indication that nodal  precession is the dominant effect, at least for a circular orbit. In the case of an eccentric orbit, an additional contribution to the TTV arises from the precession of the periastron. In principle, it can be distinguished because it produces opposite TTVs for the  transit and the occultation. 

\section{Applications}

\subsection{Systems with  measured misalignment}
At the time of writing, the projections of the obliquity of 31 transiting systems with hot Jupiters have been measured. However, this measure is usually subject to an uncertainty of $50-100$ percent when $|\lambda| \lesssim 10^\circ$. Assuming $\dot{H}=2 \times 10^{-8}$ as the detection limit, there are eight known misaligned systems for which the TDV  can be measurable. In Table~\ref{tab1}, we list their relevant parameters, from the left to the right: the name of the star, its  effective temperature, the eccentricity of the orbit, the argument of periastron, the V magnitude of the star, the references, the change of the duration of the transit in seconds per year $\Delta H$, the variation in the duration caused by nodal precession $\Delta H|_{\omega}$ in seconds per year, and, for eccentric systems,  the variation in the duration caused by apsidal precession $\Delta H|_{i}$, assuming $i_{\rm s} = 90^\circ$, also in seconds per year. We note that there are three more systems, i.e., CoRoT-2, WASP-3, and WASP-38, denoted by an asterisk in Table~\ref{tab1},  for which the TDVs can be measured if we assume the extreme possible values for their measured $\lambda $. We note that for HAT-P-2, $i=90^\circ$ so that $b'=0$ within the uncertainties and the nodal precession term vanishes at the present time.

A significant departure from the predicted TDV value can be interpreted as an indication of either a third body in the system \citep[cf. ][]{Miralda02} or an oblique planetary spin axis, if the system is very young. Specifically, if the tidal quality factor of the planet $Q^{\prime}_{\rm p} \sim 10^{8}-10^{9}$, as predicted by some models for coreless massive planets \citep[cf., e.g., ][]{GoodmanLackner09,Ogilvie09,PapaloizouIvanov10}, the tidal damping of the planetary obliquity takes place over timescales of the order of $10-100$~Myr that would lead to an observable additional nodal precession. This happens because an additional term is present in Eq.~(\ref{eqomedot0}) when the planetary spin is not aligned with the orbital angular momentum, with a coefficient $A_{\rm p} \sim 100\, A_{\rm s}$ that gives rise to a significant contribution to the variation in the inclination of the orbital plane. The detection of this effect would provide a lower limit to the largely unknown planetary tidal factors in the case of young systems \citep[cf., e.g., ][]{OgilvieLin04,GoodmanLackner09}. 

%%%%%%%%%%%%%%%%%%%%%%%%%%%%%%%%%%%%%%%%%%%%%%%%%%%%%%%%%%%%%%%%%
\begin{figure}[b]
\begin{center}
\includegraphics[width=0.7\columnwidth]{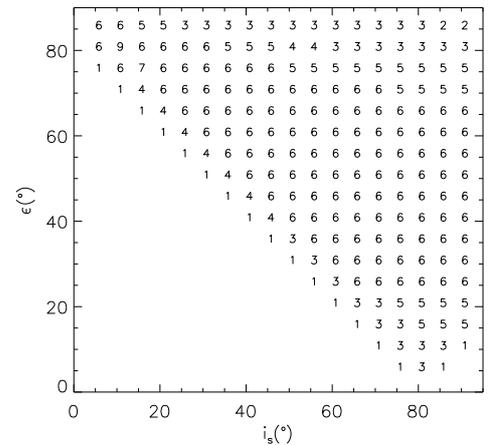}
\caption{The current number of systems with a circular orbit expected to have $\dot{H} \geq 2 \times 10^{-8}$ as a function of the obliquity $\epsilon$ and the inclination of the stellar spin to the line of sight $i_{\rm s}$, in bins of $5^\circ \times 5^\circ$. }
\label{densitymap}
\end{center}
\end{figure}
%%%%%%%%%%%%%%%%%%%%%%%%%%%%%%%%%%%%%%%%%%%%%%%%%%%%%%%%%%%%%%%%%

%%%%%%%%%%%%%%%%%%%%%%%%%%%%%%%%%%%%%%%%%%%%%%%%%%%%%%%%%%%%%%%%%%
\begin{table}[t]
\centering
\begin{tabular}{@{\extracolsep{\fill}}lcccc}
\hline\hline
Name & $T_{\rm eff}$ (K) & V mag & Refs. & $\Delta{H}$ \\
\hline
CoRoT-11     & $6440\pm120$ &$ 12.8$ & Ga10 & $-48.05$ \\
HAT-P-6     & $6570\pm80$ &$10.4$& No08 & $-1.04$ \\
HAT-P-9     & $6350\pm150$ &$12.3$& Sh09 & $-1.17$ \\
OGLE-TR-56   & $6119\pm62$ &$16.6$& To08 & $-0.67$ \\
OGLE-TR-L9   & $6933\pm60$ &$13.97 I$ & Sn09 & $-38.59$ \\
WASP-19      & $5500\pm100$ &$12.3$   & He11 & $-2.59$ \\
\hline
\end{tabular}
\caption{Transiting planetary systems with undetected misalignment and eccentricity with $\dot{H} \geq 2 \times 10^{-8}$ for $\epsilon \leq 25^\circ$. The value of $\Delta{H}$ is given in seconds per year and is computed for $i_{\rm s} = 90^\circ$ and $\epsilon=25^\circ$. References:  Ga10: \citet{Gandolfi2010}; He11: \citet{Hellier2011}; La09 \citet{Latham2009}; No08: \citet{Noyes2008};  Sh09: \citet{Shporer2009}; Sn09: \citet{Snellen2009}; To08: \citet{Torres2008}. }
\label{tab2}
\end{table}
%%%%%%%%%%%%%%%%%%%%%%%%%%%%%%%%%%%%%%%%%%%%%%%%%%%%%%%%%%%%%%%%%%%%%%%

\subsection{Systems with a circular orbit}

To date, 32 transiting exoplanetary systems with undetected eccentricity and misalignment are known. If the eccentricity of the orbit is close to zero, then the only significant source of variation of the duration of the transit is the spin-orbit misalignment. We give in Fig.~\ref{densitymap} the number of systems for with $\abs{\dot{H}} \geq 2 \times 10^{-8}$, i.e.,  expected to be detectable within three to ten years of observations, computed on a grid of $5^\circ$ resolution for $0^{\circ} \leq \epsilon \leq 85^\circ$ and $0^{\circ} \leq i_{\rm s} \leq 90^\circ$.
 The most favourable case occurs for the greatest value of the obliquity and  small values of $i_{\rm s}$, because the angular velocity of the star is higher than the projected measured value, at least for inclination values that do not lead to a rotation rate exceeding the break-up velocity. The other extreme case is more interesting, as it shows that small ob\-li\-qui\-ties can indeed be constrained by measuring TDVs. We list in Table \ref{tab2}, from left to right, the name, the effective temperature, the V magnitude, the reference for the orbit, and $\Delta{H}$ in seconds per year of the six stars for which a value $\epsilon \leq 25^\circ$ can be detected assuming a $2 \times 10^{-8}$ level of accuracy for $\dot{H}$. The three systems that have the fastest nodal precession, even for a quasi-aligned configuration, are CoRoT-11, OGLE-TR-9, and WASP-19 for which $\epsilon \leq 10^\circ$ can be measured provided that $70^\circ \leq i_{\rm s} \leq 90^\circ$.

In the case of stars hotter than $T_{\rm eff} \sim 6500$~K, an accurate measurement of $\lambda$ through the Rossiter-McLaughlin effect can be difficult owing to their rapid rotation and relative paucity of spectral lines \citep[cf., e.g., the case of CoRoT-11 in ][]{Gandolfi2010}. In those cases,  an upper limit to $\epsilon$ can be derived solely from the measurement of the nodal precession because $\sin i_{\rm s} \sin \lambda < 1$, so that Eq.~(\ref{eqomedot0}) gives
\begin{equation}
| \cos \epsilon | > \frac{1}{A_{\rm s} W_{\rm s}^{2}} \left| \langle \frac{di}{dt} \rangle \right|,
\end{equation}
where the variation in the inclination is derived from the TDV and the parameters of the transit (cf. Eq.~\ref{hdot}). 

\subsection{Systems with eccentric orbit}

In the case of eccentric systems, the duration of the transit will vary, even if the system is aligned, owing to the apsidal precession. Out of  28 eccentric systems for which no significant Rossiter-McLaughlin effect has been reported, there are three that reach $\abs{\dot{H}} \geq 2 \times 10^{-8}$. Table \ref{tab3} lists their names, the effective temperature of the star, the orbit eccentricity, the argument of the periastron, the V magnitude, the reference, the periods of apsidal precession caused by the relativistic effect $P_{\rm GR}$, the tidal bulges raised on the star and the planet  $P_{\rm tid}$, the oblateness of the star and the planet induced by their rotation $P_{\rm J_2}$, and   $\Delta{H}$ for $i_{\rm s}=90^\circ$ in seconds per year. For all of them, the predominant part of the apsidal precession is caused by the tidal bulges, and for HAT-P-23 and WASP-12, the oblateness effect is stronger than the relativistic effect, in contrast to the assumptions of \citet{Jordan2008}. However, this does not hold true for all the eccentric systems and in general we find that $P_{\rm GR} \leq P_{\rm J_2}$ and $P_{\rm GR} \approx P_{\rm tid}$.
%%%%%%%%%%%%%%%%%%%%%%%%%%%%%%%%%%%%%%%%%%%%%%%%%%%%%%%%%%%%%%%%%%%%%%%%%
\begin{table}[h]
\centering
\begin{tabular}{lccc}
\hline\hline
Name & HAT-P-23 & HAT-P-24 & WASP-12 \\
\hline
$T_{\rm eff}$ (K) & $5905 \pm 80  $ & $6373 \pm 80$ & $6250 \pm 150 $ \\
e     & $0.106 \pm 0.044$     & $0.067 \pm 0.024$     & $0.049 \pm 0.015$\\
$\omega$ ($^\circ$)     & $118 \pm 25   $ & $197 \pm 36$ & $286 \pm 0.11 $ \\
V mag & $11.94$ & $11.82$& $11.69$\\  
Refs. & Ba10 & Ba10 & He09, Hu11\\
$P_{\rm GR}$ (yr) & $2272$ & $12047$ & $1705$ \\
$P_{\rm tid}$ (yr) & $133$ & $6407$ & $18$ \\
$P_{\rm J_2}$ (yr) & $1717$ & $50243$ & $272$ \\
$\Delta{H}$ (s/yr)& $ 15.10 $ & $ 1.26 $ & $-13.42 $\\
\hline
\end{tabular}
\caption{Eccentric transiting planetary systems with undetected misalignment with $\abs{\dot{H}} \geq 2 \times 10^{-8}$. The value of $\Delta{H}$ is computed for $i_{\rm s} = 90^\circ$. References: Ba10: \citet{Bakos2010}; He09: \citet{Hebb2009};  Hu11: \citet{Husnoo2011}. }
\label{tab3}
\end{table}
%%%%%%%%%%%%%%%%%%%%%%%%%%%%%%%%%%%%%%%%%%%%%%%%%%%%%%%%%%%%
We note that for small obliquities, the rate of change in the transit duration does not vary significantly, and only two additional systems reach the level of $\abs{\dot{H}} \geq 2 \times 10^{-8}$ for $ 10^\circ \leq \epsilon \leq 25^\circ$, namely HAT-P-14/WASP-27 and HAT-P-21. We conclude that the measurement of TDVs to detect obliquity is the most interesting for systems with a circular orbit.

\section{Systems with observed TTV}
\label{observed_ttv}

The TTVs of some transiting systems have already been detected, although in some cases with an accuracy lower than considered above. The cause of the effect can be the presence of a third body in the system \citep[e.g., ][]{Miralda02,Agoletal05}, a change in the orbital period owing to the tidal interaction \citep[cf., e.g., ][]{Lanza11} or, in the case of an eccentric orbit, the apsidal precession.

Instances of TTVs have been detected for the system WASP-3 \citep{Maciejewski2010}. Even considering the highest allowed eccentricity, apsidal precession alone cannot explain the measured value. Furthermore, the period found in the $O-C$'s is three orders of magnitude shorter than the period of the precession, suggesting that the TTVs are due to a third body in the system. 

Finally, we consider the system WASP-12, for which apsidal precession, in contrast to the other one, could account for the TTVs. The eccentricity of this system is quite uncertain, and was reported to be $e = 0.049\pm0.015$ in the discovery paper of \citet{Hebb2009}, but later constrained to the much smaller value  $e=0.017^{+0.015}_{-0.011}$ \citep{Husnoo2011}. Two transits, which are 661 and 683 epochs away from the initial epoch of the ephemerides of \citet{Hebb2009}, were observed by \citet{Maciejewski2011} with a formal photometric error in the range between 0.6 and 0.7~mmag for the individual measurements. Assuming the discovery paper value for the eccentricity, we  estimate an $O-C \sim 140$~s for their epoch $683$, which is incompatible with the observed value. Adopting the eccentricity after \citet{Husnoo2011}, we find $O-C \sim 42$~s, in excellent agreement with their measured value for their epoch $683$. However, the predicted $O-C$ disagrees with that observed at their epoch $661$. Therefore,  additional high-precision photometry of this system would be required to reach a definite conclusion.

\section{Conclusions}

We have shown that the variation in the transit duration should be readily  measurable for several known  planetary systems, and can be used to constrain the value of the angle between the spin axis of the star and the orbital angular momentum. Assuming that the planet is in a quasi-synchronous and aligned state with respect to its orbit, and neglecting tidal interaction and a possible third body, we have derived the rate of change in the orbital elements produced by the deformation of the bodies resulting from the centrifugal flattening and the tidal bulges as well as the general relativity precession. After combining the results of \citet{Miralda02} and \citet{Pal2008}, we obtained a general expression for the rate of change in the transit duration, which is valid for eccentric and/or misaligned systems. Assuming an inclination of  the stellar spin axis $i_{\rm s} = 90^\circ$, a relative photometric accuracy of about $100$ ppm in a hour integration time, a typical transit depth of $0.01$~mag, and an observation interval of $3-10$ years, we found that: 
\begin{itemize}
\item eleven systems should display observable TDVs if their obliquity is consistent with the values of $\lambda$ measured through the Rossiter-McLaughlin effect; they are: CoRoT-1, CoRoT-2, HD147506/HAT-P-2, HD15082/WASP-33, Kepler-8, TrES-4, WASP-3, WASP-17, WASP-18, WASP-38, and XO-3. 
\item six systems should display observable TDVs if their eccentricity is close to zero and their obliquity smaller than $25^\circ$; they are: CoRoT-11, HAT-P-6, HAT-P-9, OGLE-TR-56, OGLE-TR-L9, and WASP-19.
\item three systems may display observable TDVs, given the uncertainty in their measured eccentricity; they are: HAT-P-23, HAT-P-24, and WASP-12; moreover, HAT-P-14/WASP-27 and HAT-P-21 could  have observable TDVs, if their obliquity is $ 10^\circ \leq \epsilon \leq 25^\circ$.
\end{itemize}

For circular systems, an upper limit to the obliquity can be obtained using TDVs, which can be particularly interesting in the case of hot host stars whose rapid rotation gives rise to a sizeable quadrupole potential and rapid nodal precession. In principle, tidal or gravitational interaction with another body and nodal precession effects can be separated  because the former two produce TTVs and TDVs, whereas the latter gives rise to  TDVs only (cf. Sect.~\ref{other_causes} for details).

The TDVs can be used to constrain the  obliquity of the star (and the planet in systems younger than $10-100$~Myr), and the $J_2$ of the star and the planet, so that they can give information about the formation scenarios, or constrain the star and planet's internal structure through the apsidal motion constant or Love number. Moreover, they can provide an indication of a third body, even of a mass comparable to that of the Earth, because this body would make an additional contribution to the nodal precession or TTV \citep[see Sects.~\ref{other_causes}, \ref{observed_ttv}, and ][]{Miralda02,Agoletal05}. As of now, the spatial mission Kepler yields the highest cumulative precision and TTVs have been used to confirm the presence of multiple planets \citep[e.g.][]{Holmanetal2010}. However, the determination of the transit durations of the non-multiple systems does  not reach the level of accuracy discussed here. In their independent analysis of Kepler-4b to Kepler-8b, \citet{Kipping2011} obtained an accuracy of several tens of seconds on the possible TDVs and TTVs for those planets using only the discovery photometry, which spans about 40 days. However, Kepler has been in operation since May 2009, meaning that the number of observed transits should have increased by a factor of about 40 at the end of the mission, leading to a significant improvement in the accuracy of the measured transit durations. In the case of Kepler-4b, a 2 $\sigma$ marginal detection of eccentricity has been claimed, but only an upper limit at $e < 0.43$ can be assumed to $95$\% confidence. Using the parameters inferred from the eccentric fit to Kepler-4b of  \citet{Kipping2011}, the apsidal precession would cause a TDV of only about $0.17^{+0.88}_{-0.58}$ s over the two years of the mission for $e=0.25$, owing to the unfavourable orientation of the orbit with respect to the line of sight ($\omega= 84.5^{+18.0}_{-18.1}$). 
Another interesting system in the Kepler field of view is HAT-P-7. However, even assuming the most favourable value of the obliquity compatible with the observations by \citet{Winnetal2009b}, we do not expect a TDV variation greater than one second over a time span of $\sim 15$~yr. Moreover, \citet{Winnetal2009b} present evidence of a third body in the system, thus a detection of a TDV by Kepler might confirm that a companion exists.

All other single planets discovered by Kepler so far have possible TDVs ranging from a few tenths of to one second over two years of observation, according to their different parameters. By the end of the mission, the accumulated photometry will surely be sufficient to place stringent constraints on the models parameters, and if the systems are misaligned, allow us to probe their stellar interiors through their TDVs.

\begin{acknowledgements}
The authors are grateful to the referee, Dr. Andras P{\'a}l , for a careful reading of the manuscript and valuable comments that helped to improve the work. 
Active star research and exoplanetary studies at INAF-Osservatorio Astrofisico di Catania and Dipartimento di Fisica e Astronomia dell'Universit\`a degli Studi di Catania 
 are funded by MIUR ({\it Ministero dell'Istruzione, dell'Universit\`a e della Ricerca}) and by {\it Regione Siciliana}, whose financial support is gratefully
acknowledged. 
This research has made use of  the ADS-CDS databases, operated at the CDS, Strasbourg, France.
\end{acknowledgements}
%%%%%%%%%%%%%%%%%%%%%%%%%%%%%%%%%%%%%%%%%%%%%%%%%%%%%%%%%%%%%%%%%
%%%%    bibliography   %%%%


\begin{thebibliography}{}
\bibliographystyle{aa}

\bibitem[Agol et al. (2005)]{Agoletal05}
Agol, E., Steffen, J., Sari, R., \& Clarkson, W.\ 2005, \mnras, 359, 567

\bibitem[Aigrain et al.(2009)]{Aigrainetal09} Aigrain, S., et al.\ 2009, \aap, 506, 425 

\bibitem[Alonso et al.(2008)]{Alonso2008} Alonso, R., et al.\ 2008, \aap, 482, L21

\bibitem[Alonso et al.(2009)]{Alonso2009} Alonso, R., et al.\ 2009, \aap, 506, 353 

\bibitem[Anderson et al.(2010)]{Anderson2010} Anderson, D.~R., et al.\ 2010, \apj, 709, 159

\bibitem[Bakos et al.(2010)]{Bakos2010} Bakos, G.~{\'A}., et al.\ 2010, \texttt{arXiv:1008.3388}

\bibitem[Ballot et al.(2006)]{Ballotetal06} 
Ballot, J., Garc{\'{\i}}a, R.~A., \& Lambert, P.\ 2006, \mnras, 369, 1281 

\bibitem[Barge et al.(2008)]{Barge2008} Barge, P., et al.\ 2008, \aap, 482, L17

\bibitem[Barros et al.(2011)]{Barros2011} Barros, S.~C.~C., et al.\ 2011, \aap, 525, A54

\bibitem[Bouchy et al.(2008)]{Bouchy2008} Bouchy, F., et al.\ 2008, \aap, 482, L25

%\bibitem[Barker \& Ogilvie (2009)]{BarkerOgilvie09} Barker, A.~J., \& Ogilvie, G.~I.\ 2009, \mnras, 395, 2268

%\bibitem[Borkovits et al. (2003)]{Borkovitsetal03}
%Borkovits, T., {\'E}rdi, B., Forg{\'a}cs-Dajka, E., \& Kov{\'a}cs, T.\ 2003, \aap, 398, 1091

\bibitem[Claret (1995)]{Claret95} Claret, A.\ 1995, \aaps, 109, 441

\bibitem[Collier Cameron et al.(2010)]{Collier2010} Collier Cameron, A., et al.\ 2010, \mnras, 407, 507

\bibitem[Croll et al.(2011)]{Croll2011} Croll, B., et al.% Lafreniere, D., Albert, L., Jayawardhana, R., Fortney, J.~J., \& Murray, N.
\ 2011, \aj, 141, 30 

%\bibitem[Cohen et al.(2010)]{Cohenetal10} Cohen, O., Drake, J.~J,
%Kashyap, V.~L., Sokolov, I.~V., \& Gombosi, T.~I.\ 2010, \apjl, 723, L64

%\bibitem[Dunham et al.(2010)]{Dunham2010} Dunham, E.~W., et al.\ 2010, \apjl, 713, L136

\bibitem[Eggleton et al.(1998)]{Eggleton1998} Eggleton, P.~P., Kiseleva, L.~G. \& Hut, P.\ 1998, \apj, 499, 853

\bibitem[Gandolfi et al.(2010)]{Gandolfi2010} Gandolfi, D., et al.\ 2010, \aap, 524, A55

\bibitem[Gillon et al. (2010)]{Gillonetal10} Gillon, M., et al.\ 2010, \aap, 511, A3 

\bibitem[Gimenez et al. (1987)]{Gimenezetal87} Gimenez, A., Kim, C.-H., \& Nha, I.-S.\ 1987, \mnras, 224, 543

\bibitem[Goodman \& Lackner (2009)]{GoodmanLackner09} Goodman, J., \& Lackner, C.\ 2009, \apj, 696, 2054

\bibitem[Hebb et al.(2009)]{Hebb2009} Hebb, L., et al.\ 2009, \apj, 693, 1920

\bibitem[Hellier et al.(2009)]{Hellier2009} Hellier, C., et al.\ 2009, \nat, 460, 1098

\bibitem[Hellier et al.(2011)]{Hellier2011} Hellier, C., et al.
%Anderson, D.~R., Collier-Cameron, A., Miller, G.~R.~M., Queloz, D., Smalley, B.,
%Southworth, J., \& Triaud, A.~H.~M.~J.
\ 2011, \apjl, 730, L31

\bibitem[Holman et al.(2010)]{Holmanetal2010} Holman, M.~J., et al.\ 2010, Science, 330, 51 

\bibitem[Husnoo et al.(2011)]{Husnoo2011} Husnoo, N., et al.\ 2011, \mnras, 424

%\bibitem[Hut(1980)]{Hut80} Hut, P.\ 1980, \aap, 92, 167

%\bibitem[Hut (1981)]{Hut81} Hut, P.\ 1981, \aap, 99, 126

%\bibitem[Iorio(2006)]{Iorio2006} Iorio, L.\ 2006, \na, 11, 490

\bibitem[Iorio(2011)]{Iorio2011} Iorio, L.\ 2011, \mnras, 411, 167

\bibitem[Jenkins et al.(2010a)]{Jenkinsetal2010a} Jenkins, J.~M., et al.\ 2010, \apjl, 713, L120 

\bibitem[Jenkins et al.(2010b)]{Jenkins2010} Jenkins, J.~M., et al.\ 2010, \apj, 724, 1108

\bibitem[Johns-Krull et al.(2008)]{Johns2008} Johns-Krull, C.~M., et al.\ 2008, \apj, 677, 657

\bibitem[Jord{\'a}n \& Bakos(2008)]{Jordan2008} Jord{\'a}n, A., \& Bakos, G.~{\'A}.\ 2008, \apj, 685, 543

\bibitem[Kipping \& Bakos(2011)]{Kipping2011} Kipping, D., \& Bakos, G.\ 2011, \apj, 730, 50

\bibitem[Kopal(1959)]{Kopal1959} Kopal, Z.\ 1959, \textit{Close binary systems}, Ed. Wiley.

\bibitem[Kozai(1962)]{Kozai1962} Kozai , Y.\ 1962, \aj, 67, 591

\bibitem[Knutson et al. (2007)]{Knutsonetal07} Knutson, H.~A., et al.\ 2007, \nat, 447, 183

%\bibitem[Lanza(2006)]{Lanza06} Lanza, A.~F.\ 2006, \mnras, 369, 1773

%\bibitem[Lanza (2010)]{Lanza10} Lanza, A. F.\ 2010, \aap, 512, A77

\bibitem[Lanza et al.(2011)]{Lanza11} Lanza, A.~F., Damiani C. \& Gandolfi, D.\ 2011, \aap, 529, A50

%\bibitem[Lanza et al.(1998)]{Lanzaetal98} Lanza, A.~F., Rodono, M., \& Rosner, R.\ 1998, \mnras, 296, 893

\bibitem[Latham et al.(2009)]{Latham2009} Latham, D.~W., et al.\ 2009, \apj, 704, 1107

\bibitem[Leconte et al. (2010)]{Leconteetal10} Leconte, J., Chabrier, G., Baraffe, I., \& Levrard, B.\ 2010, \aap, 516, A64

%\bibitem[Li et al.(2010)]{Lietal10} Li, S.-L., Miller, N., Lin, D.~N.~C., \& Fortney, J.~J.\ 2010, \nat, 463, 1054

\bibitem[Lin et al.(1996)]{Linetal96} Lin, D.~N.~C., Bodenheimer, P., \& Richardson, D.~C.\ 1996, \nat, 380, 606

%\bibitem[Lovelace et al. (2008)]{Lovelaceetal08}
%Lovelace, R.~V.~E., Romanova, M.~M., \& Barnard, A.~W.\ 2008, \mnras, 389, 1233

\bibitem[Maciejewski et al.(2010)]{Maciejewski2010} Maciejewski, G., et al.\ 2010, \mnras, 407, 2625

\bibitem[Maciejewski et al.(2011)]{Maciejewski2011} Maciejewski, G., et al.\ %Errmann, R., Raetz, S., Seeliger, M., Spaleniak, I., \& Neuh{\"a}user, R.\
2011, \aap, 528, A65

\bibitem[Mardling \& Lin(2002)]{MardlingLin02} Mardling, R.~A., \& Lin, D.~N.~C.\ 2002, \apj, 573, 829

%\bibitem[Matsumura et al. (2008)]{Matsumuraetal08}
%Matsumura, S., Takeda, G., \& Rasio, F.~A.\ 2008, \apjl, 686, L29

\bibitem[Miralda-Escud{\'e}(2002)]{Miralda02} Miralda-Escud{\'e},
J.\ 2002, \apj, 564, 1019

\bibitem[Mosser et al.(2009)]{Mosseretal09} 
Mosser, B., et al. %Baudin, F., Lanza, A.~F., Hulot, J.~C., Catala, C., Baglin, A., \& Auvergne, M.
\ 2009, \aap, 506, 245 

\bibitem[Noyes et al.(2008)]{Noyes2008} Noyes, R.~W., et al.\ 2008, \apjl, 673, L79

\bibitem[Nymeyer et al.(2010)]{Nymeyer2010} Nymeyer, S., et al.\ 2010, Bull. %etin of the 
Am. %erican 
Astron. %omical 
Soc.%iety
, 42, 1063 


\bibitem[Ogilvie(2009)]{Ogilvie09} Ogilvie, G.~I.\ 2009, \mnras, 396, 794 

\bibitem[Ogilvie \& Lin(2004)]{OgilvieLin04} Ogilvie, G.~I., \& Lin, D.~N.~C.\ 2004, \apj, 610, 477

%\bibitem[Ogilvie \& Lin (2007)]{OgilvieLin07} Ogilvie, G.~I., \& Lin, D.~N.~C.\ 2007, \apj, 661, 1180

%\bibitem[Ohta et al. (2005)]{Ohtaetal05} Ohta, Y., Taruya, A., \& Suto, Y.\ 2005, \apj, 622, 1118

\bibitem[Papaloizou \& Ivanov (2010)]{PapaloizouIvanov10} 
Papaloizou, J.~C.~B., \& Ivanov, P.~B.\ 2010, \mnras, 407, 1631 

\bibitem[Pont et al.(2010)]{Pont2010} Pont, F., et al.\ 2010, \mnras, 402, L1

\bibitem[P{\'a}l \& Kocsis(2008)]{Pal2008} P{\'a}l, A. \& Kocsis, B.\ 2008, \mnras, 389, 191

\bibitem[P{\'a}l et al.(2010)]{Pal2010} P{\'a}l, A., et al.\ 2010, \mnras, 401, 2665

%\bibitem[Peale \& Lee (2002)]{PealeLee02} Peale, S.~J., \& Lee, M.~H.\ 2002, Science, 298, 593

\bibitem[Ragozzine \& Wolf(2009)]{Ragozzine2009} Ragozzine, D., \& Wolf, A.~S.\ 2009, \apj, 698, 1778

\bibitem[Rasio \& Ford(1996)]{RasioandFord96} Rasio, F.~A., \& Ford, E.~B.\ 1996, Science, 274, 954

%\bibitem[Sari \& Goldreich (2004)]{SariGoldreich04} Sari, R., \& Goldreich, P.\ 2004, \apjl, 606, L77

\bibitem[Shporer et al.(2009)]{Shporer2009} Shporer, A., et al.\ 2009, \apj, 690, 1393

\bibitem[Simpson et al.(2010)]{Simpson2010} Simpson, E.~K., et al.\ 2010, \texttt{arXiv:1011.5664}

\bibitem[Sing \& L{\'o}pez-Morales(2009)]{Sing2009} Sing, D.~K., \& L{\'o}pez-Morales, M.\ 2009, \aap, 493, L31 

\bibitem[Snellen et al.(2009)]{Snellen2009} Snellen, I.~A.~G., et al.\ 2009, \aap, 497, 545

\bibitem[Sozzetti et al.(2009)]{Sozzetti2009} Sozzetti, A., et al.\ 2009, \apj, 691, 1145

%\bibitem[Takeda \& Rasio (2005)]{TakedaRasio05} Takeda, G., \& Rasio, F.~A.\ 2005, \apj, 627, 1001

\bibitem[Torres et al.(2008)]{Torres2008} Torres, G., Winn, J.~N., \& Holman, M.~J.\ 2008, \apj, 677, 1324

\bibitem[Triaud et al.(2010)]{Triaud2010} Triaud, A.~H.~M.~J., et al.% Cameron, A.~C., Queloz, D., Anderson, D.~R., Hellier, C., Maxted, P.~F.~L.. Mayor, M., Pepe, F., Pollacco, D., Smalley, B., West, R.~G. and Wheatley, P.~J.
\ 2010, \aap, 524, A25

\bibitem[Tripathi et al.(2010)]{Tripathi2010} Tripathi, A., et al.\ 2010, \apj, 715, 421

%\bibitem[Vidotto et al.(2010)]{Vidottoetal10} Vidotto, A.~A., Opher, M., Jatenco-Pereira, V., \& Gombosi, T.~I.\ 2010, \apj, 720, 1262

%\bibitem[Watson \& Marsh (2010)]{WatsonMarsh10} Watson, C.~A., \& Marsh, T.~R.\ 2010, \mnras, 405, 2037

\bibitem[Winn et al.(2007)]{Win2007} Winn, J.~N., et al.\ 2007, \apjl, 665, L167

\bibitem[Winn et al.(2009a)]{Win2009} Winn, J.~N., et al.\ 2009a, \apj, 700, 302

\bibitem[Winn et al.(2009b)]{Winnetal2009b} Winn, J.~N., et al.%Johnson, J.~A., Albrecht, S., Howard, A.~W., Marcy, G.~W., Crossfield, I.~J., \& Holman, M.~J.\ 
\ 2009b, \apjl, 703, L99 



%\bibitem[Winn et al.(2009)]{Winnetal09} Winn, J.~N., Holman, M.~J., Carter, J.~A., Torres, G., Osip, D.~J.,
%\& Beatty, T.\ 2009, \aj, 137, 3826

\bibitem[Winn et al.(2010)]{Winnetal10}
Winn, J.~N., et al.%, Fabrycky, D.,  Albrecht, S., \& Johnson, J.~A.
\ 2010, \apjl, 718, L145

%\bibitem[Wolff \& Simon (1997)]{WolffSimon97} Wolff, S., \& Simon, T.\ 1997, \pasp, 109, 759

%\bibitem[Zahn(2008)]{Zahn08} Zahn, J.-P.\ 2008, EAS Publications Series, 29, 67

\end{thebibliography}
\end{document}